\documentclass{emulateapj}
\shorttitle{The $M_{\rm bh}$--$M_{\rm spheroid}$ relation}
\shortauthors{Alister W.\ Graham}

\begin{document}

\title{Breaking the law: the $M_{\rm bh}$--$M_{\rm spheroid}$ relations
for core-S\'ersic and S\'ersic galaxies}

\author{Alister W.\ Graham\altaffilmark{1}}
\affil{Centre for Astrophysics and Supercomputing, Swinburne University
of Technology, Hawthorn, Victoria 3122, Australia.}
\altaffiltext{1}{Corresponding Author: AGraham@swin.edu.au}

\begin{abstract} 

The popular log-linear relation between supermassive black hole mass, $M_{\rm
bh}$, and the dynamical mass of the host spheroid, $M_{\rm sph}$, is
shown to require a significant correction. 
Core galaxies, typically with $M_{\rm bh}\gtrsim 2\times 10^8 M_{\odot}$ 
and thought to be formed in dry merger events, 
are shown to be well described by a linear relation for which the median black
hole mass is 0.36\% --- roughly double the old value of constancy.
Of greater significance is that $M_{\rm bh} \propto M_{\rm sph}^2$ 
among the (non-pseudobulge) lower-mass systems: specifically, 
$\log[M_{\rm bh}/M_{\odot}] = (1.92\pm0.38)\log[M_{\rm sph}/7\times10^{10}M_{\odot}] +
(8.38\pm0.17)$. 
`Classical' spheroids 
hosting a $10^6 M_{\odot}$ black hole will have $M_{\rm bh}/M_{\rm sph} \sim 0.025$\%.
These new 
relations presented herein  
(i) bring consistency to the relation $M_{\rm bh} \propto \sigma^5$ and the
fact that $L \propto \sigma^x$ with exponents of 5 and 2 for 
bright ($M_B \lesssim -20.5$ mag) and faint spheroids, respectively, 
(ii) mimic the non-(log-linear) behavior known to exist in the $M_{\rm
bh}$--(S\'ersic $n$) diagram,
(iii) necessitate the existence of a previously over-looked 
$M_{\rm bh}\propto L^{2.5}$ relation for S\'ersic (i.e.\ not core-S\'ersic) galaxies, and 
(iv) resolve past conflicts (in mass prediction) with the $M_{\rm
bh}$--$\sigma$ relation at the low-mass end.
Furthermore, the bent nature of the $M_{\rm bh}$--$M_{\rm
sph}$ relation reported here for `classical' spheroids 
will have a host of important implications that, while not addressed in this paper, relate to 
(i) galaxy/black hole formation theories, 
(ii) searches for the fundamental, rather than secondary, black hole scaling
relation, 
(iii) black hole mass predictions in other galaxies, 
(iv) alleged pseudobulge detections, 
(v) estimates of the black hole mass function and mass density based on luminosity
functions, 
(vi) predictions for space-based gravitational wave detections,  
(vii) connections with nuclear star cluster scaling relations, 
(viii) evolutionary studies over different cosmic epochs, 
(ix) comparisons and calibrations matching inactive black hole masses with 
low-mass AGN data, and more. 

\end{abstract}


\section{Introduction} 

The growth of supermassive black holes (SMBHs) is related to the growth of their
host-spheroid, as evinced by the existence of various $z=0$ scaling relations.
For example, the SMBH mass $M_{\rm bh}$ is tightly related to the spheroid's:
dynamical mass $M_{\rm sph}$ (e.g.\ Magorrian et al.\ 1998; Marconi \& Hunt
2003; H\"aring \& Rix 2004); 
stellar luminosity $L$ (McLure \& Dunlop 2002;
Marconi \& Hunt 2003; Graham 2007b; Beifiori et al.\ 2011; Sani et al.\ 2011; Vika et al.\ 2011); 
velocity dispersion $\sigma$ (Ferrarese \& Merritt 2000; Gebhardt et al.\
2000; Graham et al.\ 2011); 
and the radial concentration, i.e.\ S\'ersic index $n$, of the spheroid's stellar
distribution (Graham et al.\ 2001; Graham \& Driver 2007) --- at least when measured carefully along the
major-axis and the correct sky-subtraction is applied to high-$n$ galaxies 
(e.g.\ Blanton et al.\ 2005; Mandelbaum et al.\ 2005; Lauer et al.\ 2007). 

However, little attention has been given to the issue of (in)consistency with
pre-existing galaxy scaling relations. 
This paper highlights that at the heart of a crucial 
inconsistency is the (typically overlooked) bent nature of 
the luminosity-velocity dispersion relation (e.g.\ Davies et al.\
1983; Matkovi\'c \& Guzm\'an 2005).  By addressing this inconsistency --- 
previously noted in passing by (Bernardi et al.\ 2007; 
Graham\footnote{The penultimate sentence of Graham (2007b) contains a typo and
  should have read $L \propto M_{\rm bh}^{0.5}$.} 2007b, his Appendix~A; 
Graham \& Driver 2007, their section 3.2; 
Graham 2008b, his section~2.2.2) --- 
it is revealed that the $M_{\rm bh}$-$M_{\rm sph}$ and $M_{\rm bh}$-$L$
relations are better described by a {\it broken} power-law having two distinct slopes.
Given the log-linear $L$--$n$ relation (e.g.\ Graham \& Guzm\'an 2003, and
references therein), this result is in accord with the non-(log-linear) $M_{\rm
bh}$-$n$ relation (Graham \& Driver 2007). It also resolves the increasingly
noticed, but until now unexplained, problem that the previous log-linear
$M_{\rm bh}$-$M_{\rm sph}$ and $M_{\rm bh}$-$L$ relations over-predict SMBH
masses by an order of magnitude relative to the $M_{\rm bh}$-$\sigma$ relation
at low SMBH masses (e.g.\ G\"ultekin et al.\ 2011; Coziol et al.\ 2011).

\subsection{The rationale}

After Kormendy (2001) reported that 
classical bulges and pseudobulges follow the same black hole scaling relations, 
Graham (2007a, 2008a,b) and Hu (2008)
revealed that barred / pseudobulge galaxies {\it can} be offset
from what is a log-linear $M_{\rm bh}$--$\sigma$ relation defined by the 
non-barred and `classic' spheroids\footnote{Graham
(2008a) wrote that ``bar instabilities are believed to lead
to the formation of pseudobulges. Such evolution may have
resulted in (pseudo)bulges with an increased velocity dispersion
and luminosity but a relatively anemic SMBH''.}.
Given this observation, coupled with the {\it broken} $L$--$\sigma$ relation
for classical bulges and elliptical galaxies (see the review in Graham 2012a),
the $M_{\rm bh}$--$L$ and $M_{\rm bh}$--$M_{\rm sph}$ relations can not be
log-linear for such spheroids.

At the high-mass end where galaxies with partially-depleted cores 
--- thought to have formed from a small number of `dry'  
galaxy merger events (e.g.\ Begelman, Blandford \& Rees 1980; Faber et al.\
1997; Graham 2004; Bell et al.\ 2004) --- simple addition of (cold gas)-free 
early-type galaxies 
requires that the final SMBH mass increases in lock step with the
host spheroid mass and stellar luminosity 
(see also Peng 2007 and Jahnke \& Macci\`o 2011). 
To date, the $M_{\rm bh}$--$L$ relations 
have been dominated by luminous galaxies with
SMBH masses typically greater than $5 \times 10^7 M_{\odot}$. 
From these samples it has been 
found that $M_{\rm bh} \propto L^{1.0}$ (e.g.\ Graham 2007b). 
Similarly, the $M_{\rm bh}$--$M_{\rm sph}$ relation has also been reported to
have an exponent close to 1 when using bright massive spheroids 
(Marconi \& Hunt 2003; H\"aring \& Rix 2004), further supporting the dry
merger scenario. 
It has long been known that the luminous (``core'') galaxies follow the
luminosity-(velocity dispersion) relation $L \propto \sigma^5$ (e.g.\
Schechter 1980; Malumuth \& Kirshner 1981) and more recently the relation
$M_{\rm bh} \propto \sigma^5$ (e.g.\ Merritt \& Ferrarese 2001a; Hu 2008;
Graham et al.\ 2011), and thus one has that $M_{\rm bh} \propto L^1$ at the
high-mass end.

The hitherto ignored inconsistency arises from the observation that the fainter 
($M_B > -20.5$ mag) elliptical galaxies (not pseudobulges) 
do not obey the relation $L \propto \sigma^5$ but rather 
$L \propto \sigma^2$ (Davies et al.\ 1983; Held et al.\ 1992; de Rijcke et
al.\ 2005).  Samples that contain both bright (``core'') and faint (S\'ersic) 
elliptical galaxies
have an average slope of 4 or 3 depending on how far down the luminosity
function one probes (e.g.\ Faber \& Jackson 1976; Tonry 1981; 
de Vaucouleurs \& Olson 1982; Desroches et al.\ 2007). 
Davies et al.\ (1983) and Matkovi\'c \& Guzm\'an (2005) found that
the change in slope of the $L$--$\sigma$ relation 
occurs at $M_B \sim -20.5$ mag ($\sigma \sim 200$ km
s$^{-1}$), and coincides with the division between core galaxies and S\'ersic
galaxies (Graham \& Guzm\'an 2003; Graham et al.\ 2003; 
Trujillo et al.\ 2004; Gavazzi et al.\ 2005; Ferrarese et al.\ 2006). 
As reviewed and discussed in Graham (2011,2012a), 
this change in slope for elliptical galaxies has nothing to do with
pseudobulges in disc galaxies, nor the alleged divide between dwarf 
and ordinary elliptical galaxies at $M_B = -18$ 
mag, $\sigma \approx 100$ km s$^{-1}$ (Kormendy 1985; Kormendy et al.\ 2009).

In essence, given that (non-barred) S\'ersic galaxies follow the relation 
$M_{\rm bh} \propto \sigma^5$ and $L \propto \sigma^2$, then they must follow
the relation $M_{\rm bh} \propto L^{2.5}$.  This is much steeper than the
$M_{\rm bh} \propto L^1$ relation which is currently 
in use at both high and low-masses and brings into question the validity of the
extrapolation of the current relation defined by predominantly bright galaxies.  
It similarly brings into question the applicability of the log-linear relation
between the SMBH mass and the stellar mass of the host spheroid and 
the dynamical mass within the effective radius.
Given that $M_{\rm bh} \propto \sigma^5$ and 
$L\propto \sigma^2 \, \Leftrightarrow \, M_{\rm sph} \propto \sigma^2
(M/L)_{\rm dyn}$, and knowing that 
$(M/L)_{\rm dyn} \propto L^{1/4}$ (e.g.\ Faber et al.\ 1987)\footnote{Based
on the $\sigma^2 R_{\rm e}$ mass-estimate.}, 
one has the expectation that $M_{\rm bh} \propto M_{\rm sph}^2$. 

The implications and consequences of these bent black hole scaling relations
are many, and some of these are briefly discussed in section~\ref{Sec_Disc}. 
In the following section we introduce the data set that is used in
section~\ref{sec_Anal} to determine the slopes in the $M_{\rm bh}$--$M_{\rm sph}$
diagram for both ``core'' galaxies and ``S\'ersic'' galaxies, thought to have formed
form dissipationless and dissipational processes, respectively, and thus check for consistency among
the established galaxy scaling relations.

\section{Data}\label{sec_data} 

The useful H\"aring \& Rix (2004) $M_{\rm bh}$ and $M_{\rm sph}$ data set of 30
galaxies has been used in this study, with the following updates.

When available, the latest SMBH masses have been used (see the compilations in
Graham 2008b and Graham et al.\ 2011). The distances from Tonry et al.\
(2001), and thus the SMBH masses, have been reduced by 2.8\% following the
0.06 mag correction to the distance moduli, as explained in Blakeslee et al.\
(2002, their Section~4.6).

\begin{figure}
\includegraphics[angle=-90,scale=0.35]{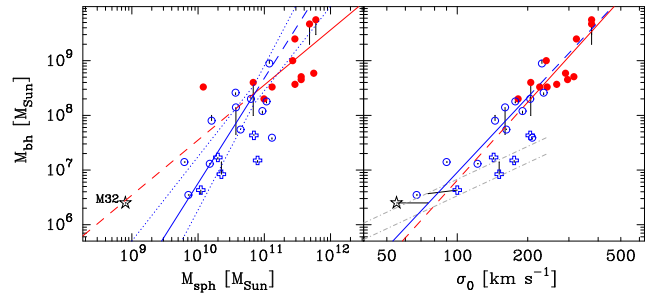}
\caption{
Optimal $M_{\rm bh}$--$M_{\rm sph}$ and $M_{\rm bh}$--$\sigma_0$ relations for
a dozen ``core'' galaxies (red dots) and a dozen non-barred S\'ersic galaxies
(blue circles).  Blue crosses denote 5 barred S\'ersic galaxies used in
Table~\ref{Tab1}.  The dashed lines show the extrapolation of these relations
beyond $M_{\rm sph}=7\times 10^{10} M_{\odot}$ The dotted lines in the left
panel delineate the 1-sigma uncertainty on the $M_{\rm bh}$--$M_{\rm sph}$
relation for the S\'ersic galaxies.
The short lines 
emanating from the data points show their old location based on the data in
H\"aring \& Rix (2004).  
The faint dot-dashed gray lines in the right panel 
correspond to a sphere-of-influence of $0.1\arcsec$ at distances 
of 3 (lower) and 6 (upper) Mpc, respectively. 
}
\label{Fig1}
\end{figure}

The velocity dispersion for the Milky Way was increased from 75 to 100 km
s$^{-1}$ (Merritt \& Ferrarese 2001a) while the velocity dispersion for 
M32 was reduced from 75 to 55 km s$^{-1}$ to reflect that of the host bulge 
(e.g.\ Lucey et al.\ 1997; I.Chilingarian 2012, in
prep.). However, with this latter update the so-called `compact elliptical' M32
(Graham 2002) 
appears to be a rather dramatic outlier from the $M_{\rm
bh}$--$M_{\rm sph}$ relation defined by the other ordinary (non-dwarf) 
S\'ersic spheroids 
and it is therefore excluded 
from the following linear regression.  Figure~\ref{Fig1} shows the original
and the new location of each galaxy in the $M_{\rm bh}$--$\sigma$ and 
$M_{\rm bh}$--$M_{\rm sph}$ diagrams. 

The breakdown of the remaining 29 galaxy types is that 12 are ``core''
galaxies, 12 are non-barred S\'ersic galaxies and 5 are barred S\'ersic
galaxies (Dullo \& Graham 2012, in preparation). 
%
%
The S\'ersic galaxy which resides, in Figure~\ref{Fig1}, 
within the region of the parameter space
where the core galaxies are found is NGC~3115 (Byun et al.\ 1996, their
figure~3).
The non-barred galaxy which resides within the region of the parameter space
where the barred galaxies are found is NGC~821. 

\section{Analysis and Results}\label{sec_Anal}

To avoid a solution which is dependent on the (somewhat disputed) measurement
errors, the regression analysis SLOPES from Feigelson \& Babu (1992) has been
used.  For the ``core'' galaxies, a symmetrical ordinary least squares (OLS)
bisector regression was used\footnote{H\"aring \& Rix (2004) employed the BCES
(Akritas \& Bershady 1996) bisector regression which factors in the adopted 
measurement errors.}.  Due to the SMBH sample selection bias --- discussed
immediately below --- which excludes data at the low-mass end,
an OLS regression of the abscissa $X$ on the ordinate
$Y$ was used for the S\'ersic galaxies.  The results
are shown in Table~\ref{Tab1}.

\begin{table}
\caption{Black hole scaling relations} 
\label{Tab1}
\begin{tabular}{@{}lccc@{}}
\hline
Galaxy Type         &     $\alpha$    &   $\beta$  &  $\Delta$ dex \\
\hline
\multicolumn{4}{c}{$\log [M_{\rm bh}/M_{\odot}] = \alpha + \beta \log [\sigma/200$ km s$^{-1}]$} \\
Core                &  $8.24\pm0.14$  &  $4.74\pm0.81$  & 0.28 \\
S\'ersic            &  $8.27\pm0.14$  &  $5.76\pm1.54$  & 0.52 \\
non-barred S\'ersic &  $8.33\pm0.12$  &  $4.57\pm1.10$  & 0.34 \\
\multicolumn{4}{c}{$\log [M_{\rm bh}/M_{\odot}] = \alpha + \beta \log [M_{\rm sph}/7\times10^{10}\, M_{\odot}$]} \\
Core                &  $8.40\pm0.37$  &  $1.01\pm0.52$  & 0.44 \\
S\'ersic            &  $8.33\pm0.20$  &  $2.30\pm0.47$  & 0.70 \\
non-barred S\'ersic &  $8.38\pm0.17$  &  $1.92\pm0.38$  & 0.57 \\
\hline
\end{tabular}

A symmetrical OLS bisector regression was used for the core-galaxies, while an
OLS$(X|Y)$ regression was used for the S\'ersic galaxies to compensate for the
sample selection bias at the low-mass end. The total rms scatter in the $\log
M_{\rm bh}$ direction is given by $\Delta$.

\end{table}

At the low mass end of the $M_{\rm
bh}$--$\sigma$ and $M_{\rm bh}$--$M_{\rm sph}$ diagram, SMBHs of a given mass
will not be detectable if the host spheroid's velocity dispersion $\sigma$ is
too high.  This is because the SMBH's gravitational sphere-of-influence
$r_{\rm inf} = G\, M_{\rm bh}/\sigma^2$ will be too small to resolve (Merritt
\& Ferrarese 2001b).  This sample selection bias --- which results in an
apparent absence of data points beneath the relations in Figure~\ref{Fig1} at
the low-$M_{\rm bh}$ end --- acts to reduce the fitted slope of the relations
for the S\'ersic galaxies (Batcheldor 2010; Graham et al.\ 2011; Schulze \&
Wisotzki 2011).  
As noted by Graham et al.\ (2011), 
while performing an OLS($X|Y$) regression helps to circumvent the problem
of the artificial floor in the $M_{\rm bh}$--$\sigma$ 
data set (see Lynden-Bell et al.\ 1988 and Feigelson
\& Babu 1992 for an understanding of this problem), 
the upwardly-sloping false-floor in the $M_{\rm bh}$--$\sigma$ and 
$M_{\rm bh}$--$M_{\rm sph}$ data means that the slopes in Table~\ref{Tab1} for
the S\'ersic galaxies will underestimate the true slope.
For galaxies at the same distance, such as
those in a cluster, lines denoting a constant size 
for the SMBHs' sphere-of-influence, such as $0.\arcsec1$, will have a slope of 2 in the $M_{\rm
bh}$--$\sigma$ diagram (see Figure~\ref{Fig1})\footnote{G\"ultekin et al.\
(2009, their section~4) mistakingly wrote that these lines have a slope of $\beta -2$,
where $\beta$ is the slope of the $M_{\rm bh}$--$\sigma$ relation.}.

There are at least three things to note when considering Figure~\ref{Fig1} and Table~\ref{Tab1}: 
(i) barred galaxies are known to be offset from the $M_{\rm bh}$--$\sigma$
relation (Graham 2007a,2008a), with their inclusion increasing the `classical'
(i.e.\ all galaxy type) slope from $\sim$5 to $\sim$6 (Graham et al.\ 2011);  
(ii) the ``core'' galaxies in Figure~\ref{Fig1} appear to have the same slope in the 
$M_{\rm bh}$--$\sigma$ diagram as the non-barred S\'ersic galaxies, and
(iii) the non-barred S\'ersic galaxies in Figure~\ref{Fig1} have a slope which is 
twice as steep as that of ``core'' galaxies in the $M_{\rm bh}$--$M_{\rm sph}$
diagram --- the significance of which can be seen in Table~\ref{Tab1}. 

For reference, the {\it total} rms scatter in the $\log M_{\rm bh}$ direction,
denoted by $\Delta$, from the H\"aring \& Rix (2004) $M_{\rm bh}$--$M_{\rm
sph}$ data about a single log-linear relation is $\sim$0.5 dex, with the value
of $\sim$0.3 dex quoted in their abstract pertaining to the intrinsic scatter.
While the lower half of our Table~\ref{Tab1} reports a total rms scatter of
0.44 dex for the core galaxies, the value of 0.57 dex for the non-barred
S\'ersic galaxies is higher --- possibly due to greater difficulties in
acquiring accurate $R_{\rm e}$ values for spheroids in disc galaxies.  This
possibility offers valid grounds for comparing the {\it intrinsic} scatter,
i.e.\ the scatter after accounting for measurement errors, but it requires
confidence in the measurement errors.  It should also be kept in mind that
our relation for the S\'ersic galaxies was constructed by minimising the
residuals in the horizontal direction rather than the $\log M_{\rm bh}$
direction.  While this level of scatter is greater than that observed for the
{\it barless} $M_{\rm bh}$--$\sigma$ relations, the sample size is small and
it would be premature to conclude which relation is more fundamental.

The log-linear relation from H\"aring \& Rix (2004), 
can substantially over-predict the SMBH masses for S\'ersic galaxies.
Relative to the second last entry in Table~\ref{Tab1}, it does so by a factor
of $\sim$5 at $M_{\rm sph}=10^{10} M_{\odot}$, and by an order of magnitude at
$M_{\rm sph}=5\times10^{9} M_{\odot}$.

For the ``core'' galaxies the $M_{\rm bh}/M_{\rm sph}$ ratio is 
roughly constant at 0.36\%, which is double the old median value of
0.14--0.2\% 
(e.g.\ Ho 1999; Kormendy 2001; Marconi \& Hunt 2003; H\"aring \& Rix 2004).
At $M_{\rm sph} = 5\times10^{11} M_{\odot}$, the new $M_{\rm bh}$--$M_{\rm 
sph}$ relation predicts SMBH masses of $1.8\times 10^9 M_{\odot}$ which is $\sim$2
times higher than the old relation's prediction of $9.6\times 10^8 M_{\odot}$.

For the non-barred S\'ersic galaxies, 
the $M_{\rm bh}/M_{\rm sph}$ mass ratio approximately varies as $M_{\rm sph}$ for
spheroids with virial masses ($\sigma^2 R_{\rm e}$) below $\sim 10^{11} M_{\odot}$.
More precisely, we have that 
$
\log \left( M_{\rm bh} / M_{\rm sph} \right) =
0.92\log \left( M_{\rm sph} / M_{\odot} \right) -12.44,
$
and it is noted that the coefficient 0.92 may be slightly underestimated due
to the sample selection bias.

\section{Discussion, Implications and Conclusions}\label{Sec_Disc}

Magorrian et al.\ (1998) wrote that there was marginal evidence for core galaxies
having a steeper dependence on $M_{\rm sph}$ than power-law\footnote{We have referred
  to ``power-law'' galaxies as S\'ersic galaxies in this paper due to their curved,
  non-(power-law) S\'ersic light profiles (Trujillo et al.\ 2004).} galaxies. 
This is the opposite behavior to what is observed here using updated SMBH masses.
The single (non-linear) log-linear relation $M_{\rm bh} \propto M_{\rm
sph}^{1.53}$ from Laor's (2001) pioneering work, (see also Salucci et al.\
2000, their figure~8, which admittedly contains potentially offset barred
spiral galaxies), can now be better understood in terms of a linear relation for
the core galaxies combined with a power-law relation for the S\'ersic galaxies.

The bent distribution seen in Figure~\ref{Fig1} can additionally be seen in
the $M_{\rm bh}$--$M_{\rm sph}$ diagram of Sani et al.\ (2011, their figure~3)
and Decarli et al.\ (2011, their figure~4) --- although it should perhaps be
noted that these authors have not actually advocated a bent relation
themselves.
Figure~\ref{Fig1} reveals a transition around $M_{\rm
  bh}$ = 1--2$\times10^8 M_{\odot}$, and 
SMBHs with masses less than this can also be 
seen to systematically reside below the single log-linear 
$M_{\rm bh}$--$L$ relation defined by the predominantly bright (``core'') galaxies
in Graham (2007b, his figure~3), 
in G\"ultekin et al.\ (2009, their figure~4), 
and in Sani et al.\ (2011, their figure~2).  
Furthermore, low-mass SMBHs are similarly offset from the relation defined
by high-mass SMBHs when their mass is plotted against the metallicity
of the host spheroid (e.g.\ Neri-Larios et al.\ 2011, their figure~3) and when
plotted against the number of globular clusters surrounding a
galaxy\footnote{A division into red and blue globular clusters may refine
this.} (Harris \& Harris 2011), revealing that both of these distributions, in
addition to the $M_{\rm bh}$--$M_{\rm sph}$ and $M_{\rm bh}$--$L$
distribution, should also be described by a broken or curved relation
rather than a single power-law.

Recent papers have tended to assume that if a galaxy has a SMBH mass that
resides beneath the single log-linear $M_{\rm bh}$--$M_{\rm sph}$ relation
defined by, for example, H\"aring \& Rix (2004), 
or beneath the single log-linear $M_{\rm bh}$--$L$ relation,
then this is evidence of a pseudobulge (e.g.\ Kormendy et al.\ 2011; Mathur et
al.\ 2011; Sani et al.\ 2011).  However this is wrong because classical
spheroids, in particular those with the lower masses and thus lower S\'ersic
indices ($\lesssim$2), reside below these old relations due to the previously
over-looked non-(log-linear) behavior of the $M_{\rm bh}$--$M_{\rm sph}$, and
$M_{\rm bh}$--$L$, distribution for classical spheroids.  These exact same
spheroids are not outliers from the $M_{\rm bh}$--$\sigma$ relation, and they
define the $L \propto \sigma^2$ relation, that is, they are not pseudobulges
(see Graham 2011 and 2012a for a fuller discussion and explanation).

Separate from the above fact, it is noted that there is some suggestion in
Figure~\ref{Fig1} that barred galaxies may be offset, to either higher
dynamical masses ($\sigma^2 R_{\rm e}$) or lower SMBH masses, from the new 
non-(log-linear) $M_{\rm bh}$--$M_{\rm sph}$ relation defined by the core and 
non-barred S\'ersic galaxies.  These offset barred galaxies may be
pseudobulges, although as Graham (2012a) details, this is difficult
to establish.  

Theories of supermassive black hole formation may require modification if they
have tied themselves to past observations of the $M_{\rm bh}$--$M_{\rm sph}$
or $M_{\rm bh}$--$L$ relation defined by massive spheroids.  This remark
extends to semi-analytical modelling, e.g.\ Croton et al.\ (2006), in which
the black hole mass growth is dominated by prescriptions set to reproduce the
$M_{\rm bh}$--$M_{\rm sph}$ relation from Marconi \& Hunt (2004) and H\"aring
\& Rix (2004).  If built by major (i.e.\ near equal mass) dry merger events,
then the one-to-one $M_{\rm bh}$--$M_{\rm sph}$ scaling relation found here is
easy to understand without any great theoretical insight.  Some focus should
additionally be spent on spheroids built through dissipational processes
involving star formation and SMBH growth, for which a linear $M_{\rm
bh}$--$M_{\rm sph}$ relation is evidently not applicable.

Past analysis of the SMBH mass function and SMBH mass density (e.g.\ Shankar
et al.\ 2004; Vika et al.\ 2009) which were based on a single log-linear
$M_{\rm bh}$--$L$ relation defined primarily by `core' galaxies will also need
to be revised.
Furthermore, past ($M_{\rm bh}$--$L$)-based predictions for the future
detection and measurement of SMBH masses in more distant galaxies, observed
with next-generation facilities, will also need to be revised.  While the
lower than expected SMBH masses in low-mass spheroids will effectively reduce
the number of detections, it may fortuitously increase the prospects for the
discovery of intermediate mass black holes ($< 10^5 M_{\odot}$).
Due to their smaller black hole masses, the potential impact of SMBH feedback
in these and slightly larger spheroids is much lower than previously thought,
and one may query whether it can in fact regulate the growth of the
surrounding spheroid --- although there is still some evidence of this in the
Milky Way (Su et al.\ 2010).

If the $L$--$\sigma$ relation turns out to be only approximated by a broken
power-law that matches some curved relation --- perhaps a log-quadratic
relation like that used for the $M_{\rm bh}$--$n$ relation (Graham \& Driver
2007) --- then a curved $M_{\rm bh}$--$M_{\rm sph}$ and $M_{\rm bh}$--$L$
relation would be preferable.  Alternatively, perhaps the $M_{\rm bh}$--$n$
relation is better described by a broken power-law.  
Studies that have failed to recover any $M_{\rm bh}$--$n$ relation appear 
to be a symptom of having 
failed to recover the well-known $L$--$n$ relation. 
This interesting topic should be possible to address through a careful
analysis in which biases on the S\'ersic index $n$ 
from unmodelled additional nuclear components,
central stellar deficits, nuclear dust or uncertainties in the point-spread
function are properly considered when using signal-to-noise weighted fitting routines
that preferentially favour the inner most point(s) of any galaxy's surface
brightness profile.  Galaxy orientation effects and ellipticity gradients can
also influence the recovered S\'ersic index. 

If $M_{\rm bh} \propto \sigma^{\beta}$ and $M_{\rm bh} \propto M_{\rm
sph}^{\gamma}$, then, using $M_{\rm sph} \propto \sigma^2\, R_{\rm e}$, 
one has that $R_{\rm e} 
\propto \sigma^{(\beta - 2\gamma)/\gamma}$.  For the core galaxies we found
that $\beta \approx 5$ and $\gamma \approx 1$, giving $R_{\rm e} \propto \sigma^3$.
Simulations of dry dissipationless mergers should follow this scaling
relation.  For the low mass spheroids ($\sigma \lesssim 200$ km s$^{-1}$) 
if $\beta \approx 5$ and
$\gamma \approx 2.0$, one has $R_{\rm e} \propto \sigma^{0.5}$. 
These predictions are best tested with a larger sample of $R_{\rm e}$ and
$\sigma$ pairs than available here. 
It is noted that above and/or below $\sigma \sim 200$ km s$^{-1}$, 
the $R_{\rm e}$--$\sigma$ relation will be 
curved if either of the above two predictor relations turn out to be curved. 
Moreover, the apparent curved nature of the $L$--$R_{\rm e}$ relation for
elliptical galaxies (Graham \& Worley 2008, their equation~16) suggests, on
the grounds of consistency, that the above two relations, or the $L$--$\sigma$
relation, may contain some curvature.  More and better data is required to
answer this question.

The curved or broken nature of the $M_{\rm bh}$--$L$ and $M_{\rm bh}$--$M_{\rm
sph}$, and $M_{\rm bh}$--$n$ (Graham \& Driver 2007), relations means that
attempts to compare the scatter about a single log-linear relation in each
diagram is an inappropriate exercise.  Claims that the log-linear $M_{\rm
bh}$--$\sigma$ relation has the least scatter of all the correlations (e.g.\
Beifiori et al.\ 2011) and is therefore the fundamental, rather than a 
secondary, relation or that the scatter in the $M_{\rm bh}$--$L$ diagram
increases at the low mass end (e.g.\ Gaskell 2011), 
must be revisited using the curved or broken relations appropriate for each data set.
Moreover, bulge+bar+disc fits are required for the barred galaxy sample, and
dust corrections are required for accurate bulge luminosities in disc
galaxies.

Due to core galaxies and S\'ersic galaxies following different relations in
the $M_{\rm bh}$--($M_{\rm sph} \sim \sigma^2 R_{\rm e})$ diagram, one may need to be
careful when trying to construct and interpret a relation to describe the
location of all galaxies on a single ($M_{\rm bh}, \sigma^2, R_{\rm e}$) plane
(e.g.\ Marconi \& Hunt 2003; Feoli \& Mele 2005, 2007; 
de Francesco et al.\ 2006; Aller \& Richstone 2007; Mancini \& Feoli 2012).  
This remark is 
additionally true when one's data contains offset barred / pseudobulge 
galaxies (see Graham 2008a).

In summary, core-galaxies and (non-barred, or non-pseudobulge) 
S\'ersic galaxies appear to follow the relation $M_{\rm bh} \propto \sigma^5$
predicted by Silk \& Rees (1998) and observed most recently by Graham et al.\
(2011) using a large updated sample with reasonable error bars on the velocity
dispersion and allowing for sample selection effects.  Dry galaxy merging 
necessitates a slope of unity for the core-galaxies' $M_{\rm bh}$--$M_{\rm
  sph}$ relation, and this is indeed observed.  Consistent with this is the
long established relation 
$L \propto \sigma^5$ for luminous (``core'') galaxies.  The fainter ($M_B > -20.5$
mag) S\'ersic galaxies follow the relation $L \propto \sigma^2$ (e.g.\ Davies et
al.\ 1983), and given that $(M/L)_{\rm dyn} \propto L^{1/4}$ (e.g.\ Faber et
al.\ 1987), one expects, and we 
find, that $M_{\rm bh} \propto M_{\rm sph}^2$ for these galaxies. 
 
Table~\ref{Tab1} reveals that although 
the $M_{\rm bh} / M_{\rm sph}$ mass ratio is constant for core galaxies
built through dry merger events (with $M_{\rm bh} \propto M_{\rm sph}^{1.0}$), 
it is not a constant value for S\'ersic galaxies.  This has important 
implications for 
research into the (mass dependent) coevolution of SMBHs and their host
galaxies when conducted by comparing local and high-$z$ black hole scaling 
relations that involve $M_{\rm sph}$ or the host's luminosity $L$ 
(e.g.\ Kisaka \& Kojima 2010; 
Lamastra et al.\ 2010; 
Schulze \& Wisotzki 2011; 
Portinari et al.\ 2011; 
Cisternas et al.\ 2011, and references therein). 
There are additionally implications for 
(i) studies which predict the SMBH mass based on the host
spheroid's mass (within $R_{\rm e}$) or luminosity, 
(ii) research on the radiative efficiency and Eddington ratios of SMBHs whose
mass is predicted using $M_{\rm sph}$ or $L$, 
(iii) cosmic event rate estimates of gravitational radiation from binary SMBHs, and
extreme mass ratio inspiral events at the centres of (nucleated) 
galaxies, when the SMBH mass is predicted using either $M_{\rm sph}$ or $L$ 
(e.g.\ Mapelli et al.\ 2011), 
(iv) studies comparing the location of low-mass AGN --- whose masses have been
obtained using reverberation mapping --- in the 
$M_{\rm bh}$--$M_{\rm sph}$ diagram against the location of the old log-linear
relation 
defined by, predominantly, high-mass black holes in non-active galaxies 
(e.g.\ Bentz et al.\ 2009; Bennert et al.\ 2011), and 
(v) connections with the (nuclear star cluster mass)-(host spheroid) relation
at the low mass-end of the $M_{\rm bh}$--$M_{\rm sph}$ relation 
(Ferrarese et al.\ 2006; Balcells et al.\ 2007; Graham \& Spitler 2009; Graham
2012b).

\section{acknowledgment}

This research was supported under the Australian Research Council’s
funding scheme (DP110103509 and FT110100263).

\end{document}